\documentclass{ws-ijmpd}

\renewcommand{\theenumi}{(\arabic{enumi})}

\newcommand{\bear}{\begin{array}}  \newcommand{\eear}{\end{array}}
\newcommand{\bea}{\begin{eqnarray}}  \newcommand{\eea}{\end{eqnarray}}
\newcommand{\beq}{\begin{equation}}  \newcommand{\eeq}{\end{equation}}
\newcommand{\bef}{\begin{figure}}  \newcommand{\eef}{\end{figure}}
\newcommand{\bec}{\begin{center}}  \newcommand{\eec}{\end{center}}
\newcommand{\rd}{{\rm d}}

\newcommand{\Eqn}[1]{&\hspace{-0.2em}#1\hspace{-0.2em}&}

\def\be{\begin{equation}}
\def\ee{\end{equation}}
\def\bea{\begin{eqnarray}}
\def\eea{\end{eqnarray}}
\def\beq{\begin{eqnarray}}
\def\eeq{\end{eqnarray}}

\def\be{\begin{equation}}
\def\ee{\end{equation}}
\def\bea{\begin{eqnarray}}
\def\eea{\end{eqnarray}}
\def\beq{\begin{eqnarray}}
\def\eeq{\end{eqnarray}}

\begin{document}

\markboth{
K. Bamba, C.Q. Geng and S. Tsujikawa
}
{
Thermodynamics in Modified Gravity Theories
}

%
\catchline{}{}{}{}{}
%

\title{ 
Thermodynamics in Modified Gravity Theories
}

\author{Kazuharu Bamba$^a$\footnote{Present address:Kobayashi-Maskawa Institute for the Origin of Particles and the Universe, 
Nagoya University, Nagoya 464-8602, Japan.}, Chao-Qiang Geng$^{a,b}$
 and Shinji Tsujikawa$^c$}

\address{$^a$Department of Physics, National Tsing Hua University,
 Hsinchu, Taiwan, R.O.C.
\\
$^b$Physics Devision, National Center for Theoretical Sciences,
 Hsinchu, Taiwan, R.O.C.\\
 $^c$Department of Physics, Faculty of Science, Tokyo University of Science\\
1-3, Kagurazaka, Shinjuku-ku, Tokyo 162-8601, Japan\\
bamba@kmi.nagoya-u.ac.jp, geng@phys.nthu.edu.tw, shinji@rs.kagu.tus.ac.jp}

\maketitle

\begin{history}
\received{Day Month Year}
\revised{Day Month Year}
\comby{Managing Editor}
\end{history}

\begin{abstract}
We demonstrate that there does exist 
an equilibrium description of thermodynamics 
on the apparent horizon in the expanding cosmological background
for a wide class of modified gravity theories with the Lagrangian 
density $f(R, \phi, X)$, where $R$ is the Ricci scalar and 
$X$ is the kinetic energy of a scalar field $\phi$. 
This comes from a suitable definition of an energy momentum tensor
of the ``dark'' component obeying the local energy conservation law 
in the Jordan frame. 
It is shown that the equilibrium description in terms of 
the horizon entropy $S$ is convenient because it 
takes into account the contribution of the horizon entropy 
$\hat{S}$ in non-equilibrium thermodynamics 
as well as an entropy production term. 
\end{abstract}

\keywords{ 
Modified theories of gravity; 
Quantum aspects of black holes, evaporation, thermodynamics; 
Dark energy; 
Cosmology.
}

\section{Introduction}
\label{I}	

The discovery of black hole entropy has implied a profound 
physical connection between gravity and thermodynamics.\cite{Beken} 
The gravitational entropy $S$ in the Einstein gravity is proportional 
to the horizon area $A$ of black holes, such that $S=A/(4G)$,
where $G$ is gravitational constant. 
A black hole with mass $M$ obeys the first law of
thermodynamics, $T \rd S=\rd M$,\cite{Bardeen} 
where $T=\kappa_s/\left(2\pi\right)$ is a Hawking temperature 
determined by the surface gravity $\kappa_s$.\cite{Hawking} 
Since black hole solutions follow from the Einstein field equations, 
the first law of black hole thermodynamics implies some 
connection between thermodynamics and the Einstein equations.
In fact, it was shown\cite{Jacobson} that the Einstein equations can 
be derived by using the Clausius relation $T\rd S=\rd Q$ on all local 
acceleration horizons in the Rindler space-time together with the relation 
$S \propto A$, where $\rd Q$ and $T$ are the energy flux across the horizon 
and the Unruh temperature seen by an accelerating observer just inside 
the horizon, respectively. 

In the theories in which the Lagrangian density $f$ is a non-linear 
function in terms of the Ricci scalar $R$ (so called ``$f(R)$ gravity''), 
it was pointed out\cite{Eling} that a non-equilibrium treatment 
is required so that the Clausius relation may be modified to 
$\rd S=\rd Q/T+\rd_i S$, where the horizon entropy $S$ is
defined by $S=F(R)A/\left(4G\right)$ with $F(R)=\partial f/\partial R$ 
and $\rd_i S$ describes a bulk viscosity entropy production term.
The variation of the quantity $F(R)$ gives rise to the non-equilibrium 
term $\rd_i S$ that is absent in the Einstein gravity. 
The reason why a non-equilibrium entropy production term 
$\rd_i \hat{S}$ appears is closely related to the theories in which 
the derivative of the Lagrangian density $f$ with respect to 
$R$ is not constant. 
It is meaningful to examine whether an equilibrium description of 
thermodynamics is possible in such modified gravity theories. 
In the present paper, we review our results\cite{Bamba:2009id} 
in Ref.~\refcite{Bamba:2009id} 
and show that equilibrium thermodynamics does exist 
for the general Lagrangian density 
$f(R, \phi, X)$ including $f(R)$ gravity and scalar-tensor 
theories, 
where $f$ is function of $R$, a scalar field $\phi$, 
and a field kinetic energy $X=-\left(\nabla \phi\right)^2/2$. 
We mention that in Ref.~\refcite{Elizalde:2008pv}, the derivation 
in Ref.~\refcite{Jacobson} was also extended to $f(R)$ gravity. 

\section{
Thermodynamics in modified gravity-non-equilibrium picture
}
\label{II}

We consider the following action
\begin{equation}
I = \frac{1}{16\pi G} \int \rd^4 x \sqrt{-g} f(R,\phi,X)
-\int \rd^4 x\,{\mathcal{L}}_{M}
(g_{\mu \nu}, \Psi_M)\,,
\label{eq:2.1}
\end{equation}
where $g$ is the determinant of the metric tensor $g_{\mu\nu}$,
${\mathcal{L}}_{M}$ is the matter Lagrangian that depends on $g_{\mu \nu}$
and matter fields $\Psi_M$, and $X=-\left(1/2\right) g^{\mu\nu}
{\nabla}_{\mu}\phi {\nabla}_{\nu}\phi$ is the kinetic term 
of a scalar field $\phi$ (${\nabla}_{\mu}$ is the 
covariant derivative operator associated with $g_{\mu \nu}$). 
The action~(\ref{eq:2.1}) can describe 
a number of 
modified gravity theories, 
e.g., $f(R)$ gravity, Brans-Dicke theories, scalar-tensor theories, 
and dilaton gravity. 
It also covers scalar field theories such as quintessence and k-essence. 
{}From the action~(\ref{eq:2.1}), the gravitational field equation and 
the equation of motion for $\phi$ are derived as
\begin{eqnarray}
& &F G_{\mu\nu}=
8\pi G T_{\mu \nu}^{(M)}+\frac12 g_{\mu \nu}
(f-RF)+\nabla_{\mu}\nabla_{\nu}F 
-g_{\mu \nu} \Box F+\frac12 f_{,X}
\nabla_{\mu} \phi \nabla_{\nu} \phi\,,
\label{eq:2.3d} \\
& & \frac{1}{\sqrt{-g}} \partial_{\mu} \left( f_{,X}
\sqrt{-g} g^{\mu \nu} \partial_{\nu} \phi \right)+f_{,\phi}=0\,,
\label{eq:2.3} 
\end{eqnarray}
where $F \equiv \partial f/\partial R$,
$f_{,X} \equiv \partial f/\partial X$,
$f_{,\phi} \equiv \partial f/\partial \phi$, 
$T_{\mu \nu}^{(M)}=(2/\sqrt{-g}) 
\delta {\cal L}_M/\delta g^{\mu \nu}$, and 
$G_{\mu\nu}=R_{\mu\nu}-\left(1/2\right)g_{\mu\nu}R$ 
is the Einstein tensor. 
Throughout this paper, we use ``${}_{,Y}$'' for the partial derivative with 
respect to the variable $Y$. 
For the matter energy momentum tensor $T^{(M)}_{\mu \nu}$, 
we consider perfect fluids of ordinary matter (radiation and non-relativistic matter) 
with total energy density $\rho_f$ and pressure $P_f$.

We assume the 4-dimensional Friedmann-Lema\^{i}tre-Robertson-Walker (FLRW) 
space-time with the metric, 
$
\rd s^2=h_{\alpha \beta} \rd x^{\alpha} \rd x^{\beta}
+\bar{r}^2 \rd \Omega^2, 
$ 
where $\bar{r}=a(t)r$ and $x^0=t, x^1=r$ with the 
2-dimensional metric $h_{\alpha \beta}={\rm diag}(-1, a^2(t)/[1-Kr^2])$.
Here, $a(t)$ is the scale factor, $K$ is the cosmic curvature, and 
$\rd \Omega^2$ is the metric of 2-dimensional sphere with unit radius. 
In the FLRW background, from Eqs.~(\ref{eq:2.3d}) and (\ref{eq:2.3}) 
we obtain the following field equations:
\begin{eqnarray}
& &3F \left( H^2+K/a^2 \right) 
=f_{,X}X+
\frac12 (FR-f)-3H\dot{F}+8\pi G \rho_f\,,
\label{fri1} \\
& & -2F \left( \dot{H}-K/a^2 \right) 
=f_{,X}X+\ddot{F}-H\dot{F}+8\pi G (\rho_f+P_f)\,,
\label{fri2}\\
& & \frac{1}{a^3} \left( a^3 \dot{\phi} f_{,X} 
\right)^{\cdot}=f_{,\phi}\,,
\label{fri3}
\end{eqnarray} 
where $H=\dot{a}/a$ is the Hubble parameter and 
the dot denotes the time derivative of $\partial/\partial t$, 
and the scalar curvature is given by 
$R=6 (2H^2+\dot{H}+K/a^2)$.
The perfect fluid satisfies the continuity equation 
$
\dot{\rho}_f+3H(\rho_f+P_f)=0.
$ 
Equations (\ref{fri1}) and (\ref{fri2}) can be written as
\begin{equation} 
H^2+\frac{K}{a^2}=\frac{8\pi G}{3F} 
\left( \hat{\rho}_d+\rho_f \right)\,, 
\quad 
\dot{H}-\frac{K}{a^2}=-\frac{4\pi G}{F}
\left( \hat{\rho}_d+\hat{P}_d+\rho_f+P_f \right)\,,
\label{frire2} 
\end{equation}
where 
\begin{eqnarray}
& & \hat{\rho}_d \equiv \frac{1}{8\pi G} \left[ f_{,X}X
+\frac12 (FR-f)-3H \dot{F} \right]\,,
\label{rodef1} \\
& & \hat{P}_d \equiv \frac{1}{8\pi G}
\left[ \ddot{F}+2H \dot{F}-\frac12 (FR-f) \right]\,.
\label{pdef1}
\end{eqnarray} 
Note that 
a hat to represent quantities in the non-equilibrium 
description of thermodynamics. 
If the density $\hat{\rho}_d$ and the pressure $\hat{P}_d$
of ``dark'' components are defined in this way, these obey the 
following equation 
\begin{equation}
\dot{\hat{\rho}}_d+3H(\hat{\rho}_d+\hat{P}_d)
=\frac{3}{8\pi G} (H^2+K/a^2) \dot{F}\,,
\label{rhocon1}
\end{equation}
where we have used Eq.~(\ref{fri3}).
For the theories with $\dot{F} \neq 0$, the right-hand side (r.h.s.) 
of Eq.~(\ref{rhocon1}) 
does not vanish, so that the standard continuity equation does not hold. 
This occurs for $f(R)$ gravity and scalar-tensor theory.

We examine the thermodynamic property of the theories given above. 
To begin with, 
the apparent horizon is determined by the condition 
$h^{\alpha \beta} \partial_{\alpha} \bar{r} \partial_{\beta} \bar{r}=0$, 
which means that the vector $\nabla \bar{r}$ is null on the surface
of the apparent horizon. For the FLRW space-time, the radius of 
the apparent horizon is given by 
$\bar{r}_A=\left( H^2+K/a^2 \right)^{-1/2}$. 
In the Einstein gravity, 
the Bekenstein-Hawking horizon entropy 
is given by $S=A/(4G)$, where $A=4\pi \bar{r}_A^2$ is the
area of the apparent horizon.\cite{Beken,Bardeen,Hawking} 
In the context of modified gravity theories, 
a horizon entropy $\hat{S}$ associated with a Noether 
charge was introduced by Wald.\cite{Wald1,Wald2} 
The Wald entropy $\hat{S}$ is a local quantity defined 
in terms of quantities on the bifurcate Killing horizon. 
More specifically, it depends on the variation of the Lagrangian 
density of gravitational theories with respect to the Riemann tensor. 
This is equivalent to 
$\hat{S}=A/(4G_{\rm eff})$, where $G_{\rm eff}=G/F$
is the effective gravitational coupling.\cite{Brustein} 
By using the Wald entropy 
\begin{equation}
\hat{S}=\frac{AF}{4G}\,,
\label{Sdef}
\end{equation}
we obtain
\begin{equation}
\frac{1}{2\pi \bar{r}_A} \rd \hat{S}=4\pi \bar{r}_A^3 H
\left( \hat{\rho}_d+\hat{P}_d+\rho_f+P_f \right)\rd t +
\frac{\bar{r}_A}{2G} \rd F\,.
\label{dSre}
\end{equation}
The apparent horizon has the following Hawking temperature 
\begin{equation}
T=\frac{|\kappa_s|}{2\pi}\,, 
\quad 
\kappa_s = -\frac{1}{\bar{r}_A}
\left( 1-\frac{\dot{\bar{r}}_A}{2H\bar{r}_A} \right)\,,
\label{tempe}
\end{equation}
where $\kappa_s$ is 
$
\kappa_s  
=
-\left(\bar{r}_A/2 \right) \left( \dot{H}+2H^2
+K/a^2 \right) 
=
-2\pi G/\left(3F\right) \bar{r}_A 
\left( \hat{\rho}_T-3\hat{P}_T \right), 
$ 
with $\hat{\rho}_T \equiv \hat{\rho}_d+\rho_f$ and 
$\hat{P}_T \equiv \hat{P}_d+P_f$. 
As long as the total equation of state $w_T=\hat{P}_T/\hat{\rho}_T$
satisfies $w_T \le 1/3$, we have $\kappa_s \le 0$, which is the 
case for standard cosmology. 
Then, Eq.~(\ref{dSre}) can be written as 
\begin{equation}
T \rd \hat{S}=4\pi \bar{r}_A^3 H (\hat{\rho}_d+
\hat{P}_d+\rho_f+P_f)\rd t 
-2\pi  \bar{r}_A^2 (\hat{\rho}_d+\hat{P}_d+\rho_f+P_f)\rd 
\bar{r}_A+\frac{T}{G}\pi \bar{r}_A^2 \rd F\,. 
\label{TdS}
\end{equation}

In the Einstein gravity, the Misner-Sharp energy\cite{Misner} is 
defined to be $E=\bar{r}_A/(2G)$. 
In $f(R)$ gravity and scalar-tensor theory, this may be extended 
to the form $\hat{E}=\bar{r}_AF/(2G)$.\cite{Gong} 
Using this latter expression for $f(R, \phi, X)$ theories, 
we find 
\begin{equation}
\hat{E}=\frac{\bar{r}_A F}{2G}=
V \frac{3F (H^2+K/a^2)}{8\pi G}=V(\hat{\rho}_d+\rho_f)\,,
\label{Edef}
\end{equation}
where $V=4\pi \bar{r}_A^3/3$ is the volume inside 
the apparent horizon. 
It follow from Eq.~(\ref{Edef}) that 
\begin{equation}
\rd \hat{E} = -4\pi \bar{r}_A^3 H (\hat{\rho}_d
+\hat{P}_d+\rho_f+P_f)\rd t 
+4\pi \bar{r}_A^2 (\hat{\rho}_d+\rho_f) 
\rd \bar{r}_A+\frac{\bar{r}_A }{2G} \rd F\,.
\label{dE}
\end{equation}
The combination of Eqs.~(\ref{TdS}) and (\ref{dE}) gives 
\begin{equation}
T \rd \hat{S}=-\rd \hat{E}+\hat{W} \rd V
+\frac{\bar{r}_A}{2G} \left( 1
+2\pi \bar{r}_A T \right) \rd F\,, 
\end{equation}
where we have introduced the work density 
defined by\cite{Hayward-1,Hayward-2,H-M-A} 
$
\hat{W} = (\hat{\rho}_d+\rho_f-\hat{P}_d-P_f)/2. 
$ 
This equation can be written in the following form: 
\begin{equation}
T \rd \hat{S}+T\rd_i \hat{S}=-\rd \hat{E}+\hat{W} \rd V\,,
\label{noneqfirst}
\end{equation}
where 
\begin{eqnarray}
\rd_i \hat{S} \Eqn{=} -\frac{1}{T} \frac{\bar{r}_A}{2G}
\left( 1+2\pi \bar{r}_A T \right) \rd F
=-\left( \frac{\hat{E}}{T}+\hat{S} \right) \frac{\rd F}{F} 
\nonumber \\ 
\Eqn{=} 
-\frac{\pi}{G} \frac{4H^2+\dot{H}+3K/a^2}
{\left(H^2+K/a^2\right)\left(2H^2+\dot{H}+K/a^2\right)}\rd F\,, 
\label{diS} 
\end{eqnarray} 
which is consistent with the result\cite{Wu2} of Ref.~\refcite{Wu2} 
for $K=0$ 
obtained in $f(R)$ gravity and scalar-tensor theories. 
The new term $\rd_i \hat{S}$ can be interpreted as a term of 
entropy production in the non-equilibrium thermodynamics. 
The theories with $F={\rm constant}$ lead to $\rd_i \hat{S}=0$, 
which means that the first-law of the equilibrium 
thermodynamics holds. 
On the other hand, 
the theories with $\rd F \neq 0$, 
including $f(R)$ gravity and scalar-tensor theory, give rise to 
the additional term (\ref{diS}).

\section{
Equilibrium description of thermodynamics in modified gravity
}
\label{III}

The energy density $\hat{\rho}_d$ 
and the pressure $\hat{P}_d$ defined in Eqs.~(\ref{rodef1}) and 
(\ref{pdef1}) do not satisfy the 
standard continuity equation for $\dot{F} \neq 0$. 
If it is possible to define $\hat{\rho}_d$ and $\hat{P}_d$ 
so that they can satisfy the conserved equation, 
the non-equilibrium description of 
thermodynamics may not be necessary. 
In this section, we demonstrate that such a treatment is indeed possible. 

One can write Eqs.~(\ref{fri1}) and (\ref{fri2}) as follows: 
\begin{equation}
3\left( H^2+\frac{K}{a^2} \right)=
8\pi G \left( \rho_d+\rho_f \right)\,, 
\quad 
-2\left( \dot{H}-\frac{K}{a^2} \right)
=8\pi G \left( \rho_d+P_d+\rho_f+P_f \right),
\label{frire2d} 
\end{equation}
where 
\begin{eqnarray}
\rho_d \Eqn{\equiv} \frac{1}{8\pi G} \biggl[ f_{,X}X
+\frac12 (FR-f)-3H \dot{F} 
+3(1-F) (H^2+K/a^2) \biggr]\,,
\label{rodef1d}
\\
P_d \Eqn{\equiv} \frac{1}{8\pi G}
\biggl[ \ddot{F}+2H \dot{F}-\frac12 (FR-f) 
-(1-F) (2\dot{H}+3H^2+K/a^2) \biggr]\,.
\label{pdef1d}
\end{eqnarray} 
If we define $\rho_d$ and $P_d$ in this way, they obey the 
following continuity equation 
$
\dot{\rho}_d+3H (\rho_d+P_d)=0,
$
where we have used Eq.~(\ref{fri3}). 
In the equilibrium description, 
the energy-momentum 
conservation in terms of ``dark'' components is met. 
Since the perfect fluid of ordinary matter also satisfies the continuity 
equation, the total energy density 
$\rho_T \equiv \rho_d+\rho_f$ 
and the total pressure $P_T \equiv P_d+P_f$ of the universe obey
the continuity equation 
$
\dot{\rho}_T + 3H (\rho_T+P_T)=0. 
$ 
Hence, the equilibrium treatment of thermodynamics can 
be executed similarly to that in the Einstein gravity. 
We introduce the Bekenstein-Hawking 
entropy\cite{Beken,Bardeen,Hawking} 
$
S=A/\left(4G\right)=\pi/\left[ G\left( H^2+K/a^2 \right) \right], 
$ 
unlike the Wald entropy. 
This allows us to obtain the equilibrium description of 
thermodynamics as that in the Einstein gravity. 
It follows that 
\begin{equation}
\frac{1}{2\pi \bar{r}_A} \rd S=4\pi \bar{r}_A^3 H
\left( \rho_d+P_d+\rho_f+P_f \right)\rd t \,.
\label{dSre2}
\end{equation}
Using the horizon temperature in Eq.~(\ref{tempe}), 
we get 
\begin{equation}
T \rd S = 4\pi \bar{r}_A^3 H (\rho_d+P_d+\rho_f+P_f)\rd t 
-2\pi  \bar{r}_A^2 (\rho_d+P_d+\rho_f+P_f)
\rd \bar{r}_A\,.
\label{TdS2}
\end{equation}
By defining the Misner-Sharp energy to be 
$ 
E=\bar{r}_A/\left(2G\right) = V\left(\rho_d+\rho_f\right), 
$
we obtain 
\begin{equation}
\rd E=-4\pi \bar{r}_A^3 H (\rho_d+P_d+\rho_f+P_f)\rd t
+4\pi \bar{r}_A^2 (\rho_d+\rho_f) 
\rd \bar{r}_A\,.
\label{dE2}
\end{equation}
Due to the conservation equation for ``dark'' components, 
the r.h.s. of Eq.~(\ref{dE2}) does not include an additional term 
proportional to $\rd F$. 
Combing Eqs.~(\ref{TdS2}) and (\ref{dE2}) gives
\begin{equation}
T \rd S=-\rd E+W \rd V\,,
\label{first}
\end{equation}
where the work density $W$ is defined by 
$
W=\left( \rho_d+\rho_f-P_d-P_f \right)/2.
$ 
Equation (\ref{first}) corresponds to the first law of 
equilibrium thermodynamics. This shows that 
the equilibrium form of thermodynamics can be derived 
by introducing the energy density $\rho_d$ and the pressure 
$P_d$ in a suitable way. 
It follows from Eq.~(\ref{first}) that 
\begin{equation}
T \dot{S}=V \left( 3H-\frac{\dot{V}}{2V} \right)
(\rho_d+\rho_f+P_d+P_f)\,.
\end{equation}
Using $V=4\pi \bar{r}_A^3/3$ and Eq.~(\ref{tempe}), 
we acquire 
\begin{equation}
\dot{S} = 
6\pi H V \bar{r}_A(\rho_d+\rho_f+P_d+P_f) 
= -\frac{2\pi}{G} \frac{H (\dot{H}-K/a^2)}
{(H^2+K/a^2)^2}\,.
\label{dSre3}
\end{equation}
The horizon entropy increases as long as the null energy condition 
$\rho_T+P_T = \rho_d+\rho_f+P_d+P_f \ge 0$ is met. 

The above equilibrium description of thermodynamics is intimately 
related with the fact that there exists an energy momentum tensor 
$T_{\mu \nu}^{(d)}$ satisfying the local conservation law 
$\nabla^{\mu} T_{\mu \nu}^{(d)}=0$.
This corresponds to writing the Einstein equation in the form 
\begin{equation}
G_{\mu \nu}=8\pi G 
\left( T_{\mu \nu}^{(d)}+
T_{\mu \nu}^{(M)} \right)\,,
\label{redeein}
\end{equation}
where 
\begin{equation}
T_{\mu \nu}^{(d)} \equiv \frac{1}{8\pi G}
\biggl[ \frac12 g_{\mu \nu}
(f-R)+\nabla_{\mu}\nabla_{\nu}F
-g_{\mu \nu} \Box F 
+\frac12 f_{,X}
\nabla_{\mu} \phi \nabla_{\nu} \phi+
(1-F)R_{\mu \nu} \biggr]\,.
\label{eneequ}
\end{equation}
Defining $T_{\mu \nu}^{(d)}$ in this way, 
the local conservation of $T_{\mu \nu}^{(d)}$ follows from 
Eq.~(\ref{redeein}) because of the relations 
$\nabla^{\mu}G_{\mu \nu}=0$ and
$\nabla^{\mu}T_{\mu \nu}^{(M)}=0$.

It can be shown that the horizon entropy $S$ in the equilibrium 
description has the following relation with $\hat{S}$ in the 
non-equilibrium description:
\begin{equation}
\rd S = \rd \hat{S} + \rd_i \hat{S}
+\frac{\bar{r}_A}{2GT}\rd F 
-\frac{2\pi (1-F)}{G}
\frac{H (\dot{H}-K/a^2)}{(H^2+K/a^2)^2}\,\rd t.
\label{Tre}
\end{equation}
By using the relations (\ref{diS}) and (\ref{dSre2}),
Eq.~(\ref{Tre}) is rewritten to the following form 
\begin{equation}
\rd S=\frac{1}{F} \rd \hat{S}+\frac{1}{F}
\frac{2H^2+\dot{H}+K/a^2}
{4H^2+\dot{H}+3K/a^2}\,\rd_i \hat{S}\,,
\label{usere}
\end{equation}
where 
\begin{equation}
\rd_i \hat{S}=-\frac{6\pi}{G} \frac{4H^2+\dot{H}+3K/a^2}
{H^2+K/a^2} \frac{\rd F}{R}\,.
\end{equation}
The difference appears in modified gravity theories with $\rd F \neq 0$, 
whereas $S$ is identical to $\hat{S}$ in the Einstein gravity ($F=1$). 
{}From Eq.~(\ref{usere}), we see that the change of the horizon entropy $S$ 
in the equilibrium framework involves the information of 
both $\rd \hat{S}$ and $\rd_i \hat{S}$ in the non-equilibrium framework.

As an example, we apply the formulas of the horizon entropies 
in the Jordan frame to inflation in $f(R)$ theories. 
In what follows, we assume the flat FLRW space-time ($K=0$). 
We consider the model $f(R)=R+\alpha R^n$ ($\alpha, n>0$) 
in the region $F=\rd f/\rd R=1+n \alpha R^{n-1} \gg 1$. 
The first inflation model for $n=2$ has been 
proposed by Starobinsky.\cite{Star80} 
During inflation, the approximations $|\dot{H}/H^2| \ll 1$ 
and $|\ddot{H}/(H\dot{H})| \ll 1$ can be used. 
With these approximations, 
in the absence of matter fluids 
the first equation in~(\ref{frire2d}) is reduced to 
$\dot{H}/H^2=-\beta,$
where 
$\beta \equiv (2-n)/\left[(n-1)(2n-1)\right].$ 
{}From this, we see that 
the scale factor behaves as 
$a \propto t^{1/\beta}$. 
Hence, 
for $\beta<1$, i.e. $n>(1+\sqrt{3})/2$, power-law inflation 
occurs. 
If $n=2$, one obtains $\beta=0$, so that $H$ is constant 
in the regime $F \gg 1$. The models with $n>2$ lead to 
the super-inflation characterized by $\dot{H}>0$ and 
$a \propto |t_0-t|^{-1/|\beta|}$ ($t_0$ is a constant). 
The standard inflation with decreasing $H$ occurs for $0<\beta<1$, 
i.e. $(1+\sqrt{3})/2<n<2$. In this case, the horizon entropy 
$
S=A/\left(4G\right)=\pi/\left( GH^2 \right) 
$ 
in the equilibrium framework 
grows as $S \propto H^{-2} \propto t^2$ during inflation. 
Meanwhile, the horizon entropy $\hat{S}=F(R)A/(4G)$
in the non-equilibrium framework has a dependence 
$\hat{S} \propto R^{n-1}/H^2 \propto H^{2(n-2)} \propto
t^{2(2-n)}$ in the regime $F \gg 1$.
Thus, $\hat{S}$ grows more slowly relative to $S$. 
This property can be understood from Eq.~(\ref{usere}), i.e.
\begin{equation}
\frac{\rd S}{\rd t}=\frac{1}{F} \frac{\rd \hat{S}}{\rd t}+
\frac{1}{F} \frac{2-\beta}{4-\beta} 
\frac{\rd_i \hat{S}}{\rd t}\,,
\label{dSrein}
\end{equation}
where 
\begin{equation}
\frac{\rd_i \hat{S}}{\rd t}=
\frac{12\pi \beta (4-\beta)}{G} HF_{,R}\,.
\label{hatSd}
\end{equation}
Here, $F_{,R} \equiv \rd F/\rd R$. 
For the above model, the term $F=n \alpha R^{n-1}$ 
evolves as $F \propto t^{2(1-n)}$. 
This means that $(1/F) \rd \hat{S}/\rd t \propto t$ in Eq.~(\ref{dSrein}), 
which has the same dependence as the time-derivative of $S$, i.e.
$\rd S/\rd t \propto t$.
The r.h.s. of Eq.~(\ref{hatSd}) is positive because $F_{,R}>0$ and $\beta>0$, 
so that $\rd_i \hat{S}/\rd t>0$. 
We have $\rd_i \hat{S}/\rd t \propto t^{3-2n}$ and therefore 
the last term on the r.h.s. of Eq.~(\ref{dSrein}) also grows 
in proportion to $t$.
As a consequence, $S$ evolves differently from $\hat{S}$ 
due to the presence of the term $1/F$.

\section{Summary}

We have explored thermodynamics 
on the apparent horizon with area $A$ in the expanding cosmological 
background for a wide class of modified gravity theories with 
the Lagrangian density $f(R, \phi, X)$. 
We have examined both non-equilibrium and equilibrium descriptions of 
thermodynamics. 

In a non-equilibrium description of thermodynamics, 
the energy density and the pressure of ``dark'' components are defined 
such that they cannot satisfy the standard continuity equation 
for the theories in which the quantity 
$F=\partial f/\partial R$ is not constant. 
In addition, 
the Wald's horizon entropy in the form $\hat{S}=AF/(4G)$ 
associated with a Noether charge is introduced, so that 
a non-equilibrium entropy production term $\rd_i \hat{S}$ appears. 
This non-equilibrium description of thermodynamics arises for the 
theories with $\rd F \neq 0$, which include $f(R)$ gravity and 
scalar-tensor theories. 

On the other hand, it is possible to acquire an equilibrium 
description of thermodynamics by defining the energy density 
and the pressure of ``dark'' components 
so that they can obey the standard continuity equation. 
In other words, 
this comes from a suitable definition of an energy momentum tensor 
of the ``dark'' component that respects to a local energy conservation law 
in the Jordan frame. 
In this framework, the horizon entropy $S$ in equilibrium thermodynamics is 
equal to the Bekenstein-Hawking entropy in the form $S=A/(4G)$, 
as in the Einstein gravity. 

Moreover, we have found that 
the variation of $S$ can be 
expressed in terms of $\rd \hat{S}$ in the non-equilibrium framework 
together with the entropy production term $\rd_i \hat{S}$.
It is considered that 
the equilibrium description of thermodynamics 
is useful not only to provide the General Relativistic analogue 
of the horizon entropy irrespective of gravitational theories 
but also to understand the nonequilibrium thermodynamics 
deeper in connection with the standard equilibrium framework.

\section*{Acknowledgments}

K.B. acknowledges the KEK theory exchange program 
for physicists in Taiwan and the very kind hospitality at 
KEK and Tokyo University of Science. 
S.T. thanks for the warm hospitality at National Tsing Hua
University where the present work was initiated. 
The work by K.B. and C.Q.G. is supported in part by 
the National Science Council of R.O.C. under 
Grant \#s: NSC-95-2112-M-007-059-MY3 and
NSC-98-2112-M-007-008-MY3
and National Tsing Hua University under the Boost Program and Grant \#s: 
97N2309F1 and 99N2539E1. 
S.T. thanks financial support for the Grant-in-Aid 
for Scientific Research Fund of the JSPS (No.~30318802)
and the Grant-in-Aid for Scientific Research on Innovative 
Areas (No.~21111006).


\end{document}